\documentclass[useAMS,usenatbib]{mn2e}
\usepackage{graphicx}
\usepackage{pstricks}

\title[Discriminating hadronic and quark stars through gravitational waves...]{Discriminating hadronic and quark stars through gravitational waves of fluid pulsation modes}
\author[C. V\'asquez Flores and G. Lugones]{C. V\'asquez Flores$^{1}$\thanks{E-mail: cesar.flores$@$ufabc.edu.br; german.lugones$@$ufabc.edu.br} and G. Lugones$^{1}$\footnotemark[1]\\
$^{1}$ Centro de Ci\^encias Naturais e Humanas, Universidade Federal do ABC,\\ Rua Santa Ad\'elia 166, Santo Andr\'e, SP, 09210-170, Brazil.\\}

\begin{document}

\maketitle

\begin{abstract}
One of the key aspects in the comprehension of neutron star interiors is the identification of
observables that may impose constraints on the equation of state.  At present, limits are obtained mainly through the study of the mass-radius relationship, the maximum rotational frequency and the cooling behaviour. However, since gravitational wave observatories such as Advanced LIGO and Advanced VIRGO will open a new window of observation of neutron stars in the very near future, it is crucial identifying observables that may emerge from the analysis of the gravitational wave emission of neutron stars.
To this end, we investigate non-radial oscillations of hadronic, hybrid and pure self-bound strange quark stars with maximum masses above the mass of the recently observed massive pulsars  PSR J1614-2230 and PSR J0348-0432 with  $M \approx 2 M_{\odot}$. For the hadronic equation of state we employ different parametrizations of a relativistic mean-field model and for quark matter we use the MIT bag model including the effect of strong interactions and color superconductivity.
We find that the first pressure mode for strange quark stars has a very different shape than for hadronic and hybrid stars. For strange quarks stars the frequency of the $p_1$ mode is larger than $6$ kHz and diverge at small stellar masses, but for  hadronic and hybrid stars it is in the range  $\sim 4-6$ kHz. This allows an observational identification of strange stars even if extra information such as the mass, the radius or the gravitational redshift of the object is unavailable or uncertain. Also, we find as in previous works that the frequency of the $g$-mode associated with the quark-hadron discontinuity in a hybrid star is in the range $0.4-1$ kHz for all masses. Thus, compact objects emitting gravitational waves above 6 kHz should be interpreted as strange quark stars and those emitting a signal within  $0.4-1$ kHz should be interpreted as hybrid stars. 
\end{abstract}


\section{Introduction}

The determination of the mass of the pulsars PSR J1614-2230 with $M = (1.97 \pm 0.04) M_{\odot}$ \citep{Demorest2010} 
and  PSR J0348-0432 with $M = (2.01 \pm 0.04) M_{\odot}$ \citep{Antoniadis2013} has brought the necessity to re-examine many aspects of the physics of neutron stars (NSs). In particular, these observations revived the discussions about whether compact stars are purely hadronic, may have quark-matter cores in their interior, or may be pure strange quark stars \citep{Ozel2010,Bejger2011,Weissenborn2011,Bednarek2012,Bonanno2012,Franzon2012,Jiang2012,Katayama2012,
Lenzi2012,Wei2012,Weissenborn2012,Alford2013,Chamel2013,Klahn2013,Lattimer2013,Mallick2013,Orsaria2013,Zdunik2013}.

In fact, several authors have looked along the years for features that may allow to distinguish unequivocally these stars through e.g. the analysis of the mass-radius relationship and the cooling behaviour since both aspects strongly depend on the microscopic composition. In the case of the mass-radius relationship a discrimination is not easy because the stellar radius is difficult to determine observationally. Moreover, at present, most observed compact stars have masses in a range where many models for hadronic, hybrid and strange stars overlap in the $M - R$ diagram. Cooling studies rely on the fact that the neutrino emission, the specific heat, the thermal conductivity,  and other relevant quantities  strongly depend on the microscopic composition.  However, while different models lead to different thermal evolution, there are many degrees of freedom in the problem and a univocal interpretation of observed data is difficult.

On the other hand, it is now clear that the increase in sensitivity of gravitational wave (GW) detectors such as Advanced LIGO and Advanced VIRGO  will bring within the next few years the science of gravitational radiation to a mode of regular astrophysical observation \citep{Riles2013}. Since NSs can be conspicuous emitters of gravitational radiation,  GWs of NSs will provide in the near future an important piece of information about several aspects of NS physics. In particular, transient phenomena involving the excitation of oscillation modes have long been considered as an important tool for the exploration of NS's interiors  because several oscillation modes may emit gravitational waves \citep{Andersson1998,Kokkotas1999}.
A lot of work has been carried out in the last three decades in order to describe the non-radial oscillatory properties of NSs; however, these studies employed equations of state that in most cases render maximum stellar masses below $2 M_{\odot}$. Recent observations have shifted the maximum stellar mass above $2 M_{\odot}$ and therefore it is worth re-examining the oscillation spectra because the change in the allowed equations of state may bring new ways to distinguish hadronic, hybrid and strange stars.

In this work, we study the $f$, $p$ and $g$ modes of hadronic, hybrid and pure self-bound strange quark stars  with maximum masses above 2 $M_{\odot}$ within the relativistic Cowling approximation. 
The paper is organized as follows: in Sec. II we describe the EoS used for the description of hadronic and quark matter. In Sec. III we present the equations that govern non-radial fluid oscillations of compact stars. In Sec. IV we present our results, and in Sec. V  a summary and our conclusions.

\section{Equations of state} 

\textit{Hadronic matter.} The relativistic mean-field  model is  widely used to describe hadronic matter in compact stars. 
In this paper we adopt the following standard Lagrangian for matter composed  by nucleons and electrons,
\begin{eqnarray} 
\label{baryon-lag}   
{\cal L}_H & = & \sum_{B} \bar{\psi}_{B}[\gamma_{\mu}(i\partial^{\mu}  - g_{\omega B}\omega^{\mu} - \frac{1}{2} g_{\rho B}\vec \tau . \vec \rho^{\mu})  \nonumber \\ 
& - & \left( m_{B} - g_{\sigma B}\sigma \right)]\psi_{B} + \frac{1}{2}({\partial_\mu \sigma \partial^\mu \sigma - m_{\sigma}^2 \sigma^2 } ) \nonumber \\ 
& - & \frac{1}{4} \omega_{\mu \nu}\omega^{\mu \nu}+ \frac{1}{2} m_{\omega}^2 \omega_\mu \omega^\mu - 
\frac{1}{4} \vec \rho_{\mu \nu}.\vec \rho^{\mu \nu} \nonumber \\
& + & \frac{1}{2} m_\rho^2 \vec \rho_{\mu}. \vec \rho^{\mu} -\frac{1}{3}bm_{n}(g_{\sigma}\sigma)^{3}-\frac{1}{4}c(g_{\sigma}\sigma)^{4}  \nonumber \\
& + & \sum_{L} \bar{\psi}_{L}    [ i \gamma_{\mu}  \partial^{\mu}  - m_{L} ]\psi_{L}.
\label{eq1}
\end{eqnarray}
Leptons $L$ are treated as non-interacting and baryons $B$ are coupled to the scalar meson $\sigma$, the isoscalar-vector meson $\omega_\mu$ and the isovector-vector meson $\rho_\mu$. For more details about the EoS obtained from the above Lagrangian the reader is referred to e.g. \citet{Lugones2013} and references therein. 
The five constants in the model are fitted to the bulk properties of nuclear matter  \citep{Glendenning2}.  In this work we use the parametrizations GM1 \citep{Glendenning2} and NL3 \citep{Lalazissis1997} whose coupling constants are shown in Table \ref{table1}.  At low densities we use the Baym, Pethick and Sutherland (BPS) model \citep{BPS}. 
%
\begin{table}
\caption{Coupling constants for the parametrizations GM1 and NL3 of the hadronic EoS. $M_{max}$  is the maximum mass of a pure hadronic star for matter composed by nucleons and electrons.}
 \begin{center}
  \begin{tabular}{l|ccc} \hline
      Set       &         GM1               &     NL3                \\ \hline
$m_\sigma$ [MeV]& 512                       &  508.194               \\
$m_\omega$ [MeV]& 783                       &  782.501               \\
$m_\rho$ [MeV]  & 770                       &  763                   \\ \hline  
 $g_\sigma$     &\ \ 8.91 \ \               &  \ \ 10.217  \ \    \\
 $g_\omega$     &\ \   10.61 \ \            &  \ \ 12.868      \ \   \\ 
 $g_\rho$       &\ \   8.196   \ \          &  \ \ 8.948  \ \    \\ 
  $b$           &  \ \ \ 0.002947 \ \ \     &  \ \  0.002055 \ \ \ \\
  $c$           &  \ \ \  -0.001070 \ \ \   &  \ \ \ -0.002651 \ \    \\
  $M_{max}$ [$M_{\odot}$] \ \ &\ \   2.32      \ \     \  &  \ \   2.73   \ \              \\   
  \hline 
 \end{tabular}    
 \end{center}
 \label{table1}
\end{table}

\textit{Unpaired quark matter.}  For non-color-superconducting quark matter we use the following modified bag model
\begin{equation}
\Omega_{QM} = \sum_{i=u,d,s,e}\Omega_{i} + \frac{3\mu^{4}}{4\pi^{2}}(1-a_{4}) + B,
\end{equation}
where $B$ is the bag constant and $\Omega_{i}$ is the thermodynamic potential for a free gas of $u$, $d$, $s$ quarks and electrons. The effects of gluon-mediated QCD interactions between the quarks in the Fermi sea are roughly incorporated through the parameter $a_4$, in the same way as in \citet{Alford2005} and \citet{Weissenborn2011}. With this equation of state we construct hybrid stars and strange quark stars. In the case of hybrid stars we consider that the hadron-quark interphase is a sharp discontinuity at which the pressure and the Gibbs free energy per baryon are continuous. For strange quark stars, the values of the parameters are chosen in order to fulfill the absolute stability condition \citep{fj84}; i.e. the energy per baryon for three (two) flavour quark matter is smaller (larger) than the energy per baryon of the most stable atomic nucleus.

\textit{Effect of color superconductivity:} We also consider colour flavour locked (CFL) strange stars \citep{Lugones2003,Lugones2004} made up of CFL quark matter from the center to the surface of the star.    Within the MIT bag model and to order $\Delta^{2}$, the thermodynamic potential $\Omega_{\rm CFL}$ can be expressed  as \citep{Lugones2002}
\begin{equation}
\Omega_{\rm CFL}= \Omega_{\rm free} - \frac{3}{\pi^{2}}\Delta^{2}\mu^{2} + B,
\end{equation}
being $\Omega_{\rm free}$ the thermodynamic potential of a state of unpaired
$u$, $d$ and $s$ quarks in which all them have a common Fermi momentum $\nu$, with $\nu$ chosen to minimize $\Omega_{\rm free}$:
\begin{equation}
\Omega_{\rm free}  =  \int_{0}^{\nu}  \frac{6 p^2 dp}{\pi^2}   [p - \mu]  + \int_{0}^{\nu} \frac{3 p^2 dp }{\pi^2}  [ \sqrt{p^2 + m_s^2} - \mu] .
\end{equation}
The binding energy of the diquark condensate is included in the condensation term proportional to $\Delta^{2}\mu^{2}$ where the chemical potential  $ \mu  \equiv  (\mu_u + \mu_d + \mu_s) / 3$ is related to $\nu$ through  $\nu = 2 \mu - ( \mu^2 + {m_s^2}/{3})^{1/2}$, being $m_s$ the mass of the strange quark. We consider $B$, $m_s$ and $\Delta$  as free parameters that fall inside the \textit{stability windows} presented in Fig. 2 of \citet{Lugones2002}; i.e. we always obtain self-bound \textit{strange stars} when integrating the stellar structure equations. Additionally, the parameters satisfy the stability condition $m_s^2 < 2 \mu \Delta$ given by \citet{Alford2004}.

\section{Non-radial fluid oscillations of compact stars}

The framework for studying non-radial modes within the theory of General Relativity  was depicted in the pioneering work of  \citet{ThorneCampollataro1967} and further extended by other authors \citep[and references therein]{DetweilerLindblom1985}. 
The perturbation equations are decomposed into spherical harmonics leading to two classes of oscillations according to the parity of the harmonics.  Even (or polar) oscillations produce spheroidal deformations on the fluid, while  odd (or axial) oscillations produce toroidal deformations (see e.g \citet{Kokkotas1999} and references therein).
For non-rotating stars composed of a perfect fluid, the fluid axial oscillations lead to a  zero frequency trivial solution to the perturbation equations with vanishing pressure and density variations while the space-time axial modes ($w$-modes or $w_{II}$-modes) are of non-zero frequency.
For polar oscillations the linearised field equations inside the star can be cast as a system of three wave equations; two of them corresponding  to the perturbations of the space-time and the other one to the density perturbations inside the star \citep{Kokkotas1999}. 
If the gravitational field is very weak, the two equations corresponding to the metric perturbation can be neglected while the remaining one  describes the oscillations of the fluid. This approach is known as the Cowling approximation and considerably simplifies the analysis.  This procedure  was first introduced by \citet{Cowling1941} for the study of Newtonian stars and subsequently adapted by  \citet{McDermott1983} for the investigation of relativistic stars.  
A more recent analysis shows that for typical relativistic stellar models the oscillation frequencies obtained by the complete linearised equations of general relativity and by the Cowling approximation differ by less than 20 \% for $f$-modes, around $10\%$ for $p$-modes \citep{Yoshida1997}, and less than a few percent for $g$-modes \citep{Sotani2001}. 
This justifies its wide utilization for studying, for example,  slowly and differentially rotating compact stars \citep{Stavridis2007}, rapidly rotating relativistic stars consisting of a perfect fluid obeying a polytropic equation of state (EoS) \citep{Boutloukos2007}, elastic modes of oscillation in the crust of a neutron star \citep{Samuelsson2007}, and neutron stars with internal anisotropic pressure  \citep{Doneva2012}.


In this work we employ the pulsation equations within the Cowling approximation as derived by  \citet{Sotani2011}. To obtain these equations,  fluid perturbations are decomposed into spherical harmonics $Y_{lm}(\theta,\phi)$ and a sinusoidal time dependence $\exp(i \omega t)$ with frequency $\omega$. The Lagrangian fluid displacements that represent the infinitesimal oscillatory perturbations of the star are
\begin{equation}
\xi^i = \left(e^{-\Lambda}W, -V\partial_\theta, -V\sin^{-2}\theta\partial_\phi\right)r^{-2}Y_{\ell m} e^{i \omega t},
\end{equation}
where $W$ and $V$ are functions of $r$. The pulsation equations read:
\begin{equation}\label{eq1}
W' = \frac{d\rho}{dP}\left[\omega^2 r^2e^{\Lambda-2\Phi}V + \Phi' W\right] - \ell(\ell+1)e^{\Lambda}V, 
\label{ecuacionparaW}
\end{equation}
\begin{equation} \label{eq2}
V' = 2\Phi'V-e^\Lambda\frac{W}{r^2}. 
\label{ecuacionparaV} 
\end{equation}
where primes denote the derivatives with respect to $r$ \citep[for more details see][]{Sotani2011}.
To close the system we need boundary conditions at the center ($r=0$) and at the surface ($r=R$) of the star. The behaviour of $W$ and $V$ near the center of the star can be obtained from the above equations and is given by  $W(r)=Cr^{\ell+1}+{\cal O}(r^{\ell+3})$ and $V(r) = -Cr^\ell/\ell+{\cal O}(r^{\ell+2})$,  where $C$ is an arbitrary constant. At the surface of the star the Lagrangian perturbation in the pressure must be zero ($\Delta P=0$), leading to
\begin{equation}
\omega^2 r^2e^{\Lambda-2\Phi}V + \Phi' W = 0.
\label{boundary2}
\end{equation}
In the case of hybrid stars, we have to impose additional junction conditions at the density discontinuity between the quark and the hadronic phases. These junction conditions read \citep{Sotani2011}
\begin{eqnarray}
  W_+  &=& W_-, \label{junction1}\\
  V_+ &=& \frac{e^{2\Phi}}{\omega^2 {R_g}^2}\Big( \frac{\rho_-+P}{\rho_++P} 
       \big[\omega^2{R_g}^2e^{-2\Phi}V_{-} + e^{-\Lambda}\Phi' W_-\big] \nonumber \\ 
&  & - e^{-\Lambda}\Phi' W_+\Big), \label{junction2}
\end{eqnarray}
where $R_g$ represents the position of the density discontinuity, and the values of $W$, $V$, and $\rho$ at both sides of the discontinuity are denoted by:  $W_- \equiv W(R_g-0)$, $V_-\equiv V(R_g-0)$, $\rho_- \equiv \rho(R_g-0)$, $W_+ \equiv W(R_g+0)$, $V_+\equiv V(R_g+0)$, and $\rho_+ \equiv \rho(R_g+0)$.

In order to numerically solve the oscillation equations we proceed as follows.
First, we integrate the Tolman-Oppenheimer-Volkoff stellar structure equations for each set of parameters of the equation of state  in order to obtain the coefficients
of the oscillation equations for a given central pressure.
Then we solve the oscillation equations by means of the  {\em shooting method}:
we start the numerical integration of Eqs. (\ref{ecuacionparaW}) and (\ref{ecuacionparaV}) for a
trial value of $\omega^2$ and a given set of values of
$W$ and $V$ such that the boundary condition at the centre is fulfilled.  
The equations are integrated outwards trying to
match the boundary condition at the star's surface. After each integration, the
trial value of $\omega^2$ is corrected through a Newton-Raphson iteration scheme
in order to improve the matching of the
surface boundary condition until the desired precision is achieved.
The discrete values of $\omega$ for which Eq.
(\ref{boundary2}) is satisfied are the eigenfrequencies of the star. 
In order to check our code we have reproduced the results of \citet{LindblomSplinter1990}.
In the case of hybrid stars, we employ the {\em shooting to a fitting point} method. The numerical integration is started at the centre and the surface of the star towards the density discontinuity and the trial value of $\omega^2$ is corrected until the junction conditions in Eqs. (\ref{junction1}) and (\ref{junction2}) are verified with the desired precision.

\begin{figure}
 \centering
\includegraphics[angle=0,scale=0.45]{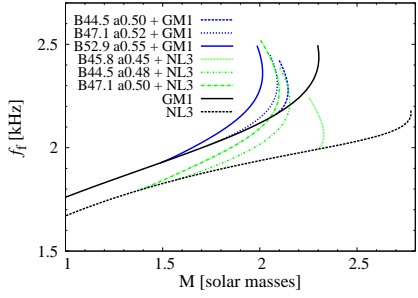}
\caption{The frequency $f_f = \omega_f /(2 \pi)$ of the $f$-mode for hadronic stars and hybrid stars as a function of the stellar mass $M$ for models with maximum masses above 2 $M_{\odot}$. For hadronic matter the GM1 and NL3 parametrizations are used. For stars containing quark matter, the labels indicate the values of $B$ in  MeV fm$^{-3}$ and of $a_4$ (label ``a").  }
\label{fig_f_hybrid}
\end{figure}

\begin{figure}
\centering
\includegraphics[angle=0,scale=0.45]{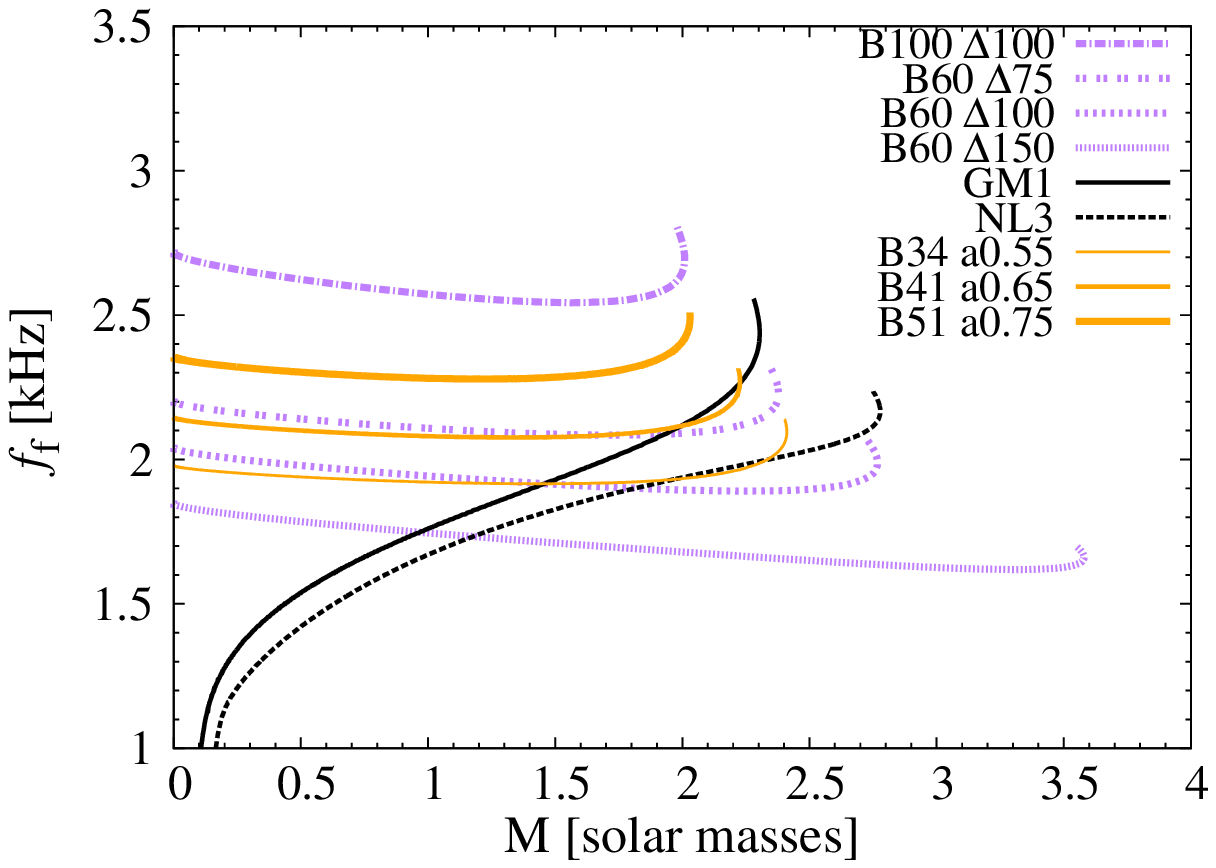}
\includegraphics[angle=0,scale=0.45]{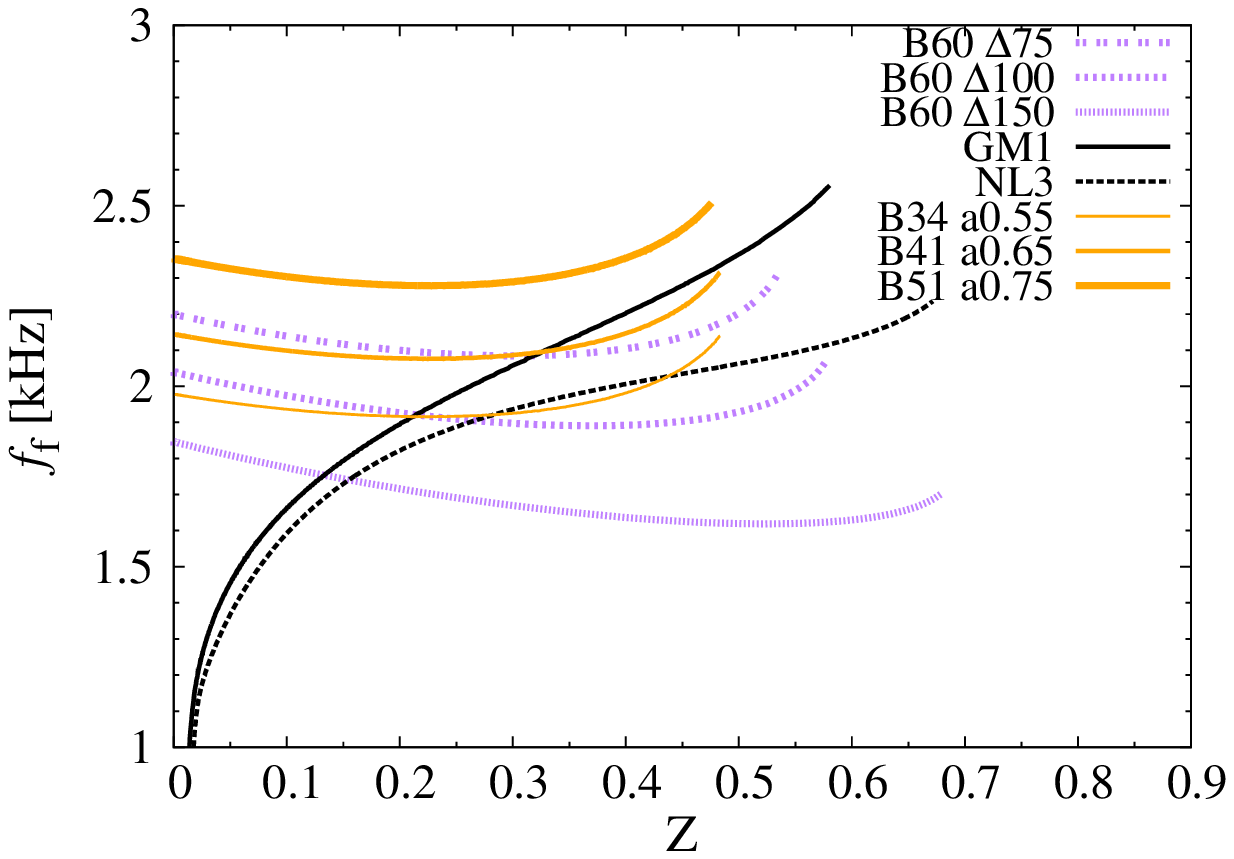}
\includegraphics[angle=0,scale=0.45]{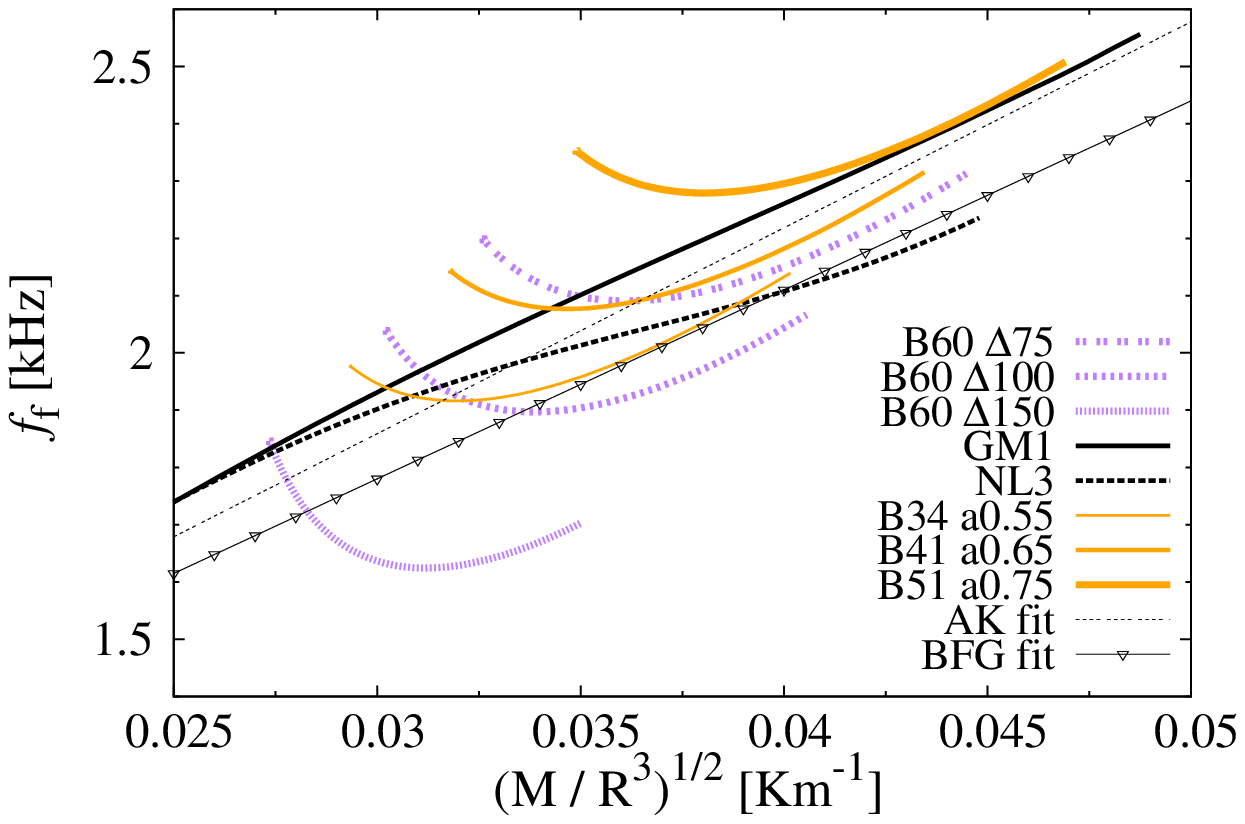}
\caption{Frequencies of the $f$-mode for hadronic stars and strange quark stars as a function of the mass $M$,  the gravitational redshift $z$ at the surface of the star, and the square root of the mean density. For strange stars, the labels indicate the values of $B$ in  MeV fm$^{-3}$, of $\Delta$ in  MeV and of $a_4$.  In the lower panel we include the analytic fittings of  \citet{Andersson1998} and  \citet{Benhar2004} for hadronic stars.}
\label{fig_fmode}
\end{figure}

\section{Results}

\begin{figure}
 \centering
\includegraphics[angle=0,scale=0.45]{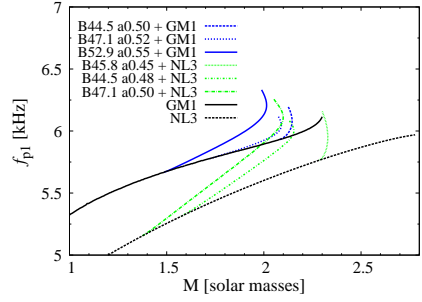}
\caption{Frequencies of the first pressure mode for models of  hadronic and hybrid stars with maximum masses above 2 $M_{\odot}$.}
\label{fig_p1_hybrid}
\end{figure}

\begin{figure}
 \centering
\includegraphics[angle=0,scale=0.45]{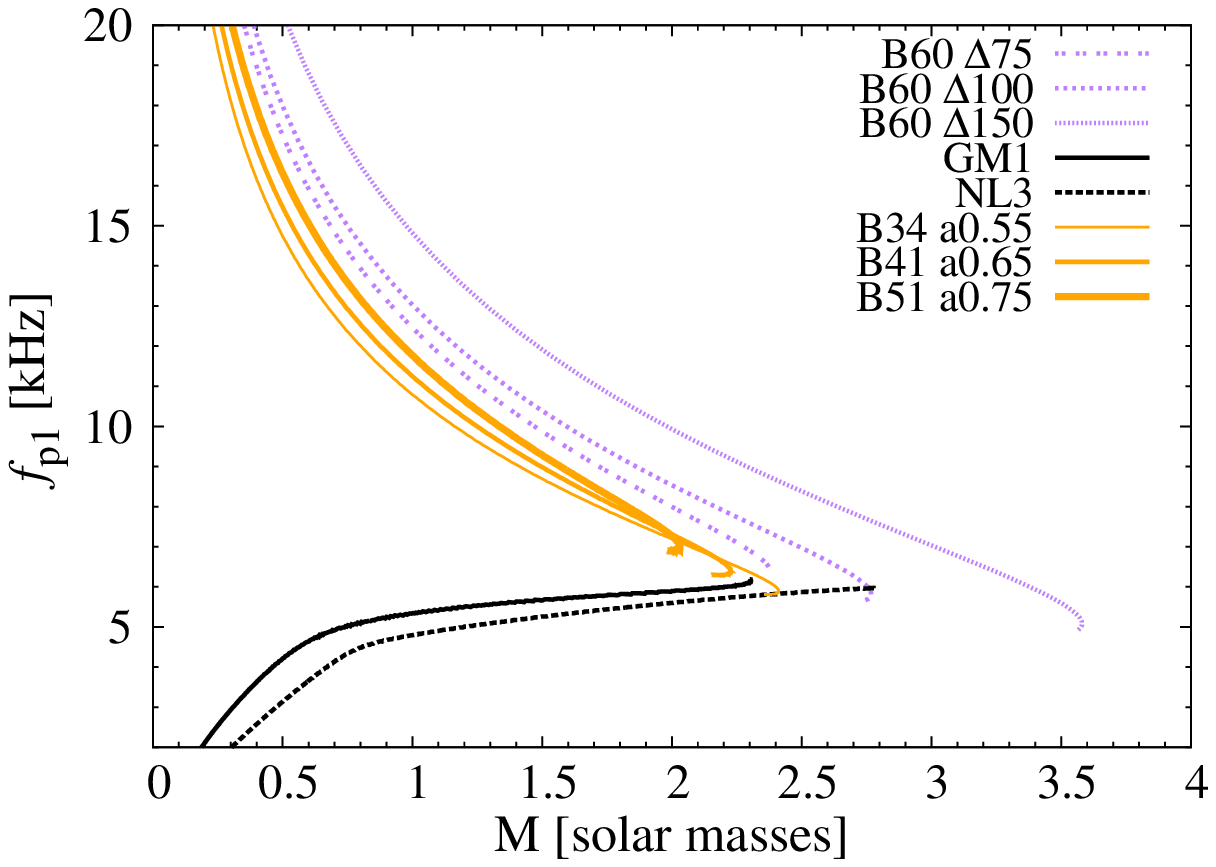} 
\includegraphics[angle=0,scale=0.45]{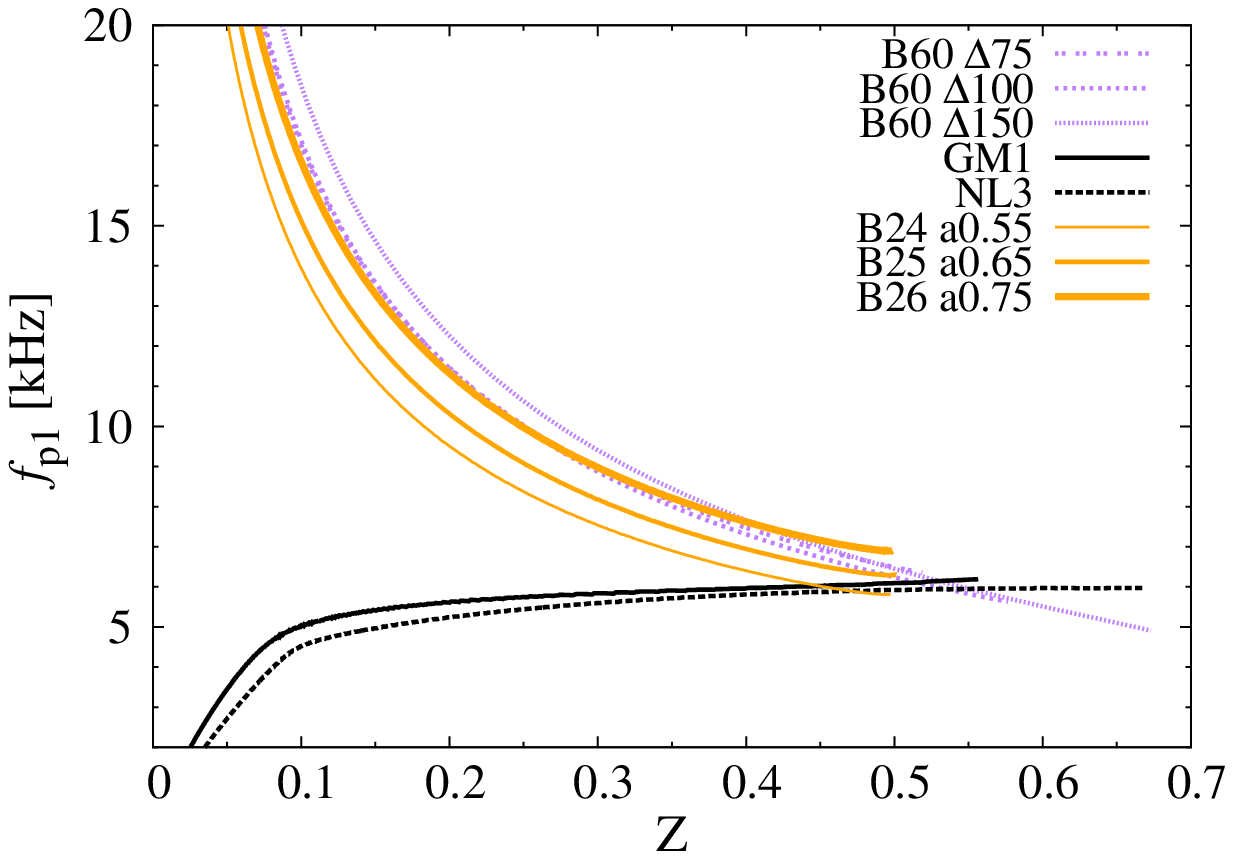} 
\caption{Frequencies of the first pressure mode for models of hadronic and strange quark stars with maximum mass above 2 $M_{\odot}$. From this figure and the previous one we note that the curves for strange stars don't overlap with the curves for hybrid or hadronic stars for the most relevant values of $M$ and $z$ ($M$ in the range $1-2 M_{\odot}$ and $z$ around $0.35$). }
\label{fig_p1mode}
\end{figure}

\begin{figure}
 \centering
\includegraphics[angle=0,scale=0.48]{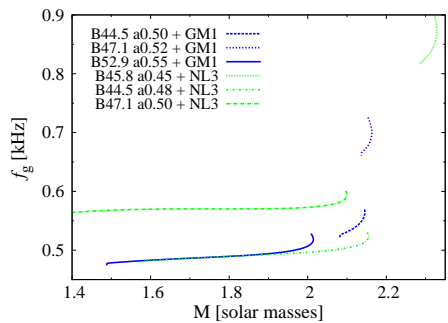}
\caption{Frequency of the $g$-mode for hybrid stars with maximum masses above $2 M_{\odot}$.
The curves fall in the range $0.4-1$ kHz and don't overlap with other modes. }
\label{fig_gmode}
\end{figure}

The polar quasi-normal modes are usually classified according to a scheme in which each family of modes is directly associated with the restoring force that prevails when a fluid element is displaced from its equilibrium position \citep{Cowling1941}. The most important modes for gravitational wave emission are the (pressure) $p$-modes, the (fundamental) $f$-mode,  and the (gravity)  $g$-modes. The frequencies of  $g$-modes are lower than those of $p$-modes, and the two sets are separated by the frequency of the $f$-mode \citep{Kokkotas1999}. These are called \textit{fluid modes} to distinguish them from e.g. purely gravitational modes ($w$-modes) for which the fluid motion is barely excited. Since the metric perturbations are set to zero within the Cowling approximation, only $f$, $p$ and $g$ modes can be studied through the equations of the previous section. In chemically homogeneous, zero temperature (and hence isentropic)  stars, all the $g$-modes are zero frequency \citep{Finn1987}, i.e. in the present study they arise only in hybrid stars.  
In Figs. \ref{fig_f_hybrid}$-$\ref{fig_gmode}  we show our results for quadrupole oscillations ($l=2$). For strange quark stars and hybrid stars, the mass of the strange quark has been set to $m_s = 100$ MeV in all calculations, and we spanned the values of the parameters $a_4$, $B$ and $\Delta$  that give stars with a maximum mass above $2 M_{\odot}$.

In Fig. \ref{fig_f_hybrid} we show our results for the $f$-mode of hadronic and hybrid stars and in Fig. \ref{fig_fmode} of hadronic and strange stars. 
In Fig. \ref{fig_f_hybrid} we see that there is a folding in the curve for hybrid stars at the mass value above which the star has a core of quark matter. Above that mass, the curves for hybrid stars are steeper than the hadronic ones but the models overlap each other.
In the upper panel of Fig. \ref{fig_fmode} we show $f_f$ as a function of the stellar mass for hadronic and strange stars. The shape of the curves is qualitatively different for both types of objects but  the results tend to overlap around $\sim 2$ kHz in the mass range of interest.  However, in some cases it is possible to differentiate strange stars from hadronic/hybrid stars. For example, objects in the mass range $1-1.5 M_{\odot}$ with $f_f$ in the range $2-3$ kHz would be strange stars.
We also present the behaviour of $f_f$ as a function of the gravitational redshift $z$ at the surface of the star (see central panel of Fig. \ref{fig_fmode}) because $z$ could be inferred through the identification of spectral lines. 
Finally, in the bottom panel of Fig. \ref{fig_fmode}  we show $f_f$ as a function of the square root of the average density, which is a more natural scaling in the case of hadronic stars \citep{Andersson1998}. We also show the fitting formulae found by \citet{Andersson1998} and \citet{Benhar2004} which are in reasonable agreement with the curves for hadronic stars with maximum masses above 2 $M_{\odot}$. 
The main conclusion that can be obtained from Figs. \ref{fig_f_hybrid} and \ref{fig_fmode} is that
there is an overlapping of the results for hadronic, hybrid and strange stars around a frequency of $\sim 2$ kHz, and therefore, in most cases it would be rather difficult to infer the internal composition of a compact object even if its mass or the surface $z$ is determined together with the frequency of the fundamental mode. However, in some cases the identification of strange stars would be possible. 

Our results for the $f$-mode of hybrid/hadronic stars are in agreement with similar calculations by \cite{Benhar2007} and \cite{Sotani2011} which give $f_f \approx 1.5-3.5$ kHz; however, notice that most of their models have maximum masses well below  $2 M_{\odot}$.  Additionally, our calculations were performed for many values of the stellar mass and therefore our curves show clearly the behaviour near the maximum mass and in the case of hybrid stars the folding  at the mass value above which the star has a core of quark matter. For strange stars,  \cite{Benhar2007} found results in agreement with ours; in particular, they show that strange stars can be differentiated from hadronic/hybrid stars in some cases.  However, their models have maximum masses that never exceed $1.8 M_{\odot}$, and thus they are incompatible with the recent observations PSR J1614-2230 and PSR J0348-0432. Notice that we also explored the effect of color superconductivity that was not addressed in previous works.

In Figs. \ref{fig_p1_hybrid}$-$\ref{fig_p1mode} we show our results for the first pressure mode. The shape of the curves for hadronic and hybrid stars is different as can be seen in Fig. \ref{fig_p1_hybrid}. Above $M \sim 1.5 M_{\odot}$, the branches corresponding to hybrid stars emerge over the curves corresponding to the hadronic models, i.e. for a given hadronic equation of state the frequencies for hybrid stars are larger than for hadronic stars. However, as seen in Fig.  \ref{fig_p1_hybrid}, the curves overlap if we consider several hadronic and hybrid models, and again, it would be rather difficult to infer the existence of a quark matter core inside a given compact star even if $M$ or $z$ are measured together with $f_{p1}$.  The situation is different if we compare strange quark stars with hybrid or hadronic stars. 
For hadronic and hybrid stars, the frequencies are $\sim 6$ kHz near the maximum mass and decrease to $\sim 4$ kHz for small masses. For strange stars, the frequencies are also $\sim 6$ kHz near the maximum mass but are considerably larger for smaller masses. Neutron stars observed up to date have masses in the range $1.0 - 2.0 ~ M_{\odot}$, i.e. in a region where the curves for strange and hadronic/hybrid stars don't overlap. Therefore,  the observation  a  $p_1$-mode with a frequency significantly larger than $\sim 6$ kHz would be a clear evidence in favour of a strange quark star, even if the mass, the radius or the gravitational redshift of the object are unknown. 

Notice that our results for the $p_1$-mode are consistent with previous studies. In \cite{Benhar2004} purely hadronic stars and hybrid stars  were described using some hadronic equations of state \citep{APR,BBS,Glendenning2} and the MIT bag model for quark matter, giving  $f_{p1} \sim 5-6$ kHz, as in our calculations. They also consider few strange star models with low maximum masses around $1.5 M_{\odot}$ and obtain $f_{p1} \sim 8-11$ kHz, in agreement with our results for low mass objects. 
\cite{Sotani2011} also present calculations for hadronic and hybrid models with very low maximum masses
that are consistent with our results.

In Fig. \ref{fig_gmode} we present the results for the  $g$-mode of hybrid stars. The frequencies cover the range $0.4-1$ kHz, in agreement with previous calculations by \cite{Miniutti2003} for polytropic equations of state and \cite{Sotani2011} for hybrid stars with low maximum masses. The frequency interval of the $g$-modes for different parametrizations of the EOS is clearly separated from the $f$-mode frequencies. Additionally, other $g$-modes such as those associated 
with a non-homogeneous composition in the outer layers of the star, or those associated with a thermal profile, have lower frequencies than the here-studied 
quark-hadron-discontinuity $g$-modes \citep{Miniutti2003}.  Thus, the observation of oscillations with frequency in the range $0.4-1$ kHz would be an evidence of a hybrid star.

\section{Summary and Conclusions}

%
\begin{table}
\centering
\begin{tabular}{l|ccc}
\hline 
            &       $f_f$        & $f_{p1}$   & $f_{g}$   \\
\hline 
strange stars  &       $\sim 2$ kHz &  $> 6$ kHz    &   not present   \\
hybrid stars   &       $\sim 2$ kHz &  $\sim 4-6$ kHz    &  $\sim  0.4-1$ kHz \\
hadronic stars &       $\sim 2$ kHz &  $\sim 4-6$ kHz    &   not present   \\
\hline
\end{tabular}
\caption{Discrimination between hadronic, hybrid and strange quark stars based on the observation of the $f$, $p_1$ and $g$ modes.} 
\label{table2}
\end{table}

In this paper we have investigated non-radial fluid oscillations of hadronic, hybrid and strange quark stars with maximum masses above the mass of the recently observed pulsars  
PSR J1614-2230 and PSR J0348-0432 with  $M \approx 2 M_{\odot}$.  For the hadronic equation of state we employed two different parametrizations of a relativistic mean-field model with nucleons and electrons. For quark matter we have included the effect of strong interactions and color superconductivity within the MIT bag model. The equations of non-radial oscillations were integrated within the Cowling approximation in order to determine the frequency of the $f$, $p_1$ and $g$-modes.

We find that the fundamental mode is sensitive to the internal composition, but due to the uncertainties in the equations of state, there is an overlapping of the curves corresponding to hadronic, hybrid and strange quark stars for stellar masses larger that $ \sim 1 ~ M_{\odot}$.  As a consequence it would be difficult to distinguish hybrid and hadronic stars through the $f$-mode frequency, even if the mass or the surface $z$ of the object is determined concomitantly with $f_f$. However, there are features that in some cases may allow a differentiation between strange stars and hadronic/hybrid stars. For example, strange stars cannot emit gravitational waves with frequency below $\sim 1.7$ kHz for any value of the mass. Also, sources with a mass in the range $1-1.5 M_{\odot}$ emitting a signal in the range $2-3$ kHz would be strange stars. 
The frequency of the $p_1$ mode is much more affected by the internal composition of the star.  For hadronic and hybrid stars, we find that $f_{p1}$ is in the range $4-6$ kHz for objects with masses in the range $1-2 \, M_{\odot}$, but for strange quark stars it is always significantly larger than $\sim 6$ kHz.  Thus, a compact object emitting a signal above $\sim 6$ kHz could be identified as a strange star even if its  mass or gravitational redshift are unknown. 
High frequency $g$-modes are only present in hybrid stars and fall in the range $0.4-1$ kHz. Thus, they are clearly distinguishable from the fundamental mode, and of low-frequency $g$-modes associated with 
chemical inhomogeneities in the outer layers or thermal profiles. Our results are summarized in Table \ref{table2} and show that based on the simultaneous analysis of the frequency of the $f$, $p_1$ and $g$-modes it would be possible to discriminate between hadronic, hybrid and strange quark stars. 

The spectrum of a pulsating compact star is very rich and therefore the possibility of having other kind of modes in the same frequency range of the above studied fluid modes should be studied in more detail in the light of modern equations of state leading to stellar models compatible with the recent observations of PSR J1614-2230 and PSR J0348-0432. If other modes are present in the same range, the criterion presented above may be less effective.  Additionally, rotation is known to change the frequency range of the modes, but these modifications are not expected to alter qualitatively the scenario presented in this paper \cite[see e.g.][and references therein]{{Gaertig2009}}. In particular, it looks quite robust the conclusion that compact objects emitting a signal above 6 kHz should be interpreted as strange quark stars and those emitting a signal in the range  $\sim 0.4-1$ kHz should be interpreted as hybrid stars.  

\section{Acknowledgements}
C. V\'asquez Flores and G. Lugones acknowledge the financial support received from  FAPESP-Brazil.


\begin{thebibliography}{}

\bibitem[\protect\citeauthoryear{Akmal et al.}{1998}]{APR} Akmal  A., Pandharipande V R., Ravenhall D. G., 1998, Phys. Rev. C, 58, 1804

\bibitem[\protect\citeauthoryear{Alford et al.}{2004}]{Alford2004}  Alford M., Kouvaris C., Rajagopal K., 2004, Phys. Rev. Lett., 92, 222001

\bibitem[\protect\citeauthoryear{Alford et al.}{2005}]{Alford2005}  Alford M.,  Braby M., Paris M.,  Reddy S., 2005, ApJ, 629, 969


\bibitem[\protect\citeauthoryear{Alford et al.}{2013}]{Alford2013} Alford M. G., Han S., Prakash  M., 2013,  arXiv:1302.4732 [astro-ph.SR]

\bibitem[\protect\citeauthoryear{Andersson \& Kokkotas}{1998}]{Andersson1998} Andersson N.,  Kokkotas K. D., 1998, MNRAS, 299, 1059

\bibitem[\protect\citeauthoryear{Antoniadis et al.}{2013}]{Antoniadis2013} Antoniadis  J.  et al., 2013, Science 340, 6131 


\bibitem[\protect\citeauthoryear{Baldo et al.}{2000}]{BBS} Baldo M., Burgio G. F., Schulze H. J., 2000, Phys. Rev. C, 61, 055801


\bibitem[\protect\citeauthoryear{Baym et al.}{1971}]{BPS} Baym G.,  Pethick C. J., Sutherland P., 1971, ApJ, 170, 299

\bibitem[\protect\citeauthoryear{Bednarek et al.}{2012}]{Bednarek2012}  Bednarek I. et al., 2012,  A\&A 543, A157 

\bibitem[\protect\citeauthoryear{Bejger et al.}{2011}]{Bejger2011} Bejger M.,  Zdunik J. L.,  Haensel P., Fortin M., 2011, A\&A 536, A92 

\bibitem[\protect\citeauthoryear{Benhar et al.}{2004}]{Benhar2004}  Benhar O., Ferrari V.,  Gualtieri L., 2004, Phys. Rev. D, 70, 124015


\bibitem[\protect\citeauthoryear{Benhar et al.}{2007}]{Benhar2007} Benhar O., Ferrari V., Gualtieri L., Marassi S., 2007, Gen. Relativ. Gravit. 39, 1323


\bibitem[\protect\citeauthoryear{Bonanno \& Sedrakian}{2012}]{Bonanno2012} Bonanno L., Sedrakian A., 2012, A\&A, 539, A16

\bibitem[\protect\citeauthoryear{Boutloukos \&  Nollert}{2007}]{Boutloukos2007}  Boutloukos S., Nollert H. P., 2007, Phys. Rev. D, 75, 43007

\bibitem[\protect\citeauthoryear{Chamel et al.}{2013}]{Chamel2013} Chamel N. et al., 2013, A\&A 553, A22

\bibitem[\protect\citeauthoryear{Cowling}{1941}]{Cowling1941}  Cowling, T. G., 1941,  MNRAS, 101, 509

\bibitem[\protect\citeauthoryear{Demorest et al.}{2010}]{Demorest2010} Demorest P. B., Pennucci T., Ransom S. M., Roberts M. S. E., Hessels J. W. T., 2010, Nature, 467, 1081

\bibitem[\protect\citeauthoryear{Detweiler \& Lindblom}{1985}]{DetweilerLindblom1985}  Detweiler S., Lindblom L., 1985, ApJ, 292, 12

\bibitem[\protect\citeauthoryear{do Carmo et al.}{2013}]{Lugones2013} do Carmo T. A. S., Lugones G.,  Grunfeld A. G., 2013, J. Phys. G: Nucl. Part. Phys., 40, 035201

\bibitem[\protect\citeauthoryear{Doneva \&  Yazadjiev}{2012}]{Doneva2012}  Doneva D. D.,  Yazadjiev S. S., 2012, Phys. Rev. D, 85, 124023

\bibitem[\protect\citeauthoryear{Farhi \& Jaffe}{1984}]{fj84}  Farhi E.,  Jaffe R. L., 1984, Phys. Rev. D, 30, 2379

\bibitem[\protect\citeauthoryear{Finn}{1987}]{Finn1987}  Finn L. S., 1987,  MNRAS, 227, 265

\bibitem[\protect\citeauthoryear{Franzon et al.}{2012}]{Franzon2012}  Franzon B., Fogaca D. A., Navarra F.S., Horvath J.E., 2012, Phys. Rev. D86, 065031

\bibitem[\protect\citeauthoryear{Gaertig \&  Kokkotas}{2009}]{Gaertig2009} Gaertig E. and  Kokkotas K. D., 2009, Phys. Rev. D, 80, 064026

\bibitem[\protect\citeauthoryear{Glendenning \& Moszkowski}{1991}]{Glendenning2} Glendenning N. K., Moszkowski S. A., 1991, Phys. Rev. Lett., 67, 2414

\bibitem[\protect\citeauthoryear{Horvath \& Lugones}{2004}]{Lugones2004}  Horvath J. E., Lugones G., 2004, A\&A, 422, L1

\bibitem[\protect\citeauthoryear{Jiang et al.}{2012}]{Jiang2012}	 Jiang W.-Z., Li B.-A.,  Chen L.-W., 2012, ApJ 756, 56 

\bibitem[\protect\citeauthoryear{Katayama et al.}{2012}]{Katayama2012}   Katayama T., Miyatsu T.,  Saito K., 2012, ApJS 203, 22 

\bibitem[\protect\citeauthoryear{Klahn et al.}{2013}]{Klahn2013} Klahn T., Lastowiecki R.,  Blaschke D. B.,  2013, arXiv:1307.6996 [nucl-th]

\bibitem[\protect\citeauthoryear{Kokkotas \& Schmidt}{1999}]{Kokkotas1999}  Kokkotas K. D., Schmidt B. G., 1999, Living Rev. Relativity, 2, 2

\bibitem[\protect\citeauthoryear{Lattimer \& Lim}{2013}]{Lattimer2013} 	Lattimer J. M.,  Lim Y., 2013, ApJ 771, 51

\bibitem[\protect\citeauthoryear{Lalazissis et al.}{1997}]{Lalazissis1997} Lalazissis G. A., K\"onig J., Ring P., 1997, Phys. Rev. C, 55, 540

\bibitem[\protect\citeauthoryear{Lindblom \& Splinter}{1990}]{LindblomSplinter1990} Lindblom L.,  Splinter R. J., 1990, ApJ, 348, 192

\bibitem[\protect\citeauthoryear{Lugones \& Horvath}{2002}]{Lugones2002}  Lugones G.,  Horvath J. E., 2002, Phys. Rev. D, 66, 074107

\bibitem[\protect\citeauthoryear{Lugones \& Horvath}{2003}]{Lugones2003}  Lugones G.,  Horvath J. E., 2003, A\&A, 403, 173

\bibitem[\protect\citeauthoryear{Lenzi \& Lugones}{2012}]{Lenzi2012}  Lenzi  C. H., Lugones  G., 2012, ApJ 759, 57

\bibitem[\protect\citeauthoryear{Mallick}{2013}]{Mallick2013}  Mallick  R., 2013, Phys.Rev. C87,  025804

\bibitem[\protect\citeauthoryear{McDermott et al.}{1983}]{McDermott1983}  McDermott P. N., Van Horn H. M.,  Scholl J. F., 1983, ApJ, 268, 837

\bibitem[\protect\citeauthoryear{Miniutti et al.}{2003}]{Miniutti2003} Miniutti G.,  Pons J. A., Berti E., Gualtieri L., Ferrari  V., 2003, MNRAS 338, 389

\bibitem[\protect\citeauthoryear{Orsaria et al.}{2013}]{Orsaria2013}  Orsaria M. et al., 2013, Phys.Rev. D87,  023001

\bibitem[\protect\citeauthoryear{Ozel et al.}{2010}]{Ozel2010}  Ozel F. , Psaltis D., Ransom S., Demorest P., Alford  M., 2010, ApJ 724, L199 

\bibitem[\protect\citeauthoryear{Riles}{2013}]{Riles2013} Riles K., 2013, Prog. Part. Nucl. Phys. 68, 1

\bibitem[\protect\citeauthoryear{Samuelsson \& Andersson}{2007}]{Samuelsson2007} Samuelsson L.,  Andersson N., 2007,  MNRAS, 374, 256 

\bibitem[\protect\citeauthoryear{Sotani et al.}{2001}]{Sotani2001}  Sotani H., Tominaga K., Maeda  K.I., 2001, Phys. Rev. D,  65, 024010

\bibitem[\protect\citeauthoryear{Sotani et al.}{2004}]{Sotani2004}  Sotani H., Kohri K., Harada T., 2004, Phys. Rev. D, 69, 84008

\bibitem[\protect\citeauthoryear{Sotani et al.}{2011}]{Sotani2011} Sotani H., Yasutake N., Maruyama T.,  Tatsumi T., 2011, Phys. Rev. D, 83, 024014

\bibitem[\protect\citeauthoryear{Stavridis et al.}{2007}]{Stavridis2007}  Stavridis A.,  Passamonti A.,  Kokkotas K., 2007, Phys. Rev. D, 75, 64019

\bibitem[\protect\citeauthoryear{Thorne \& Campollataro}{1967}]{ThorneCampollataro1967} Thorne K. S., Campolattaro A., 1967, ApJ, 149, 591

\bibitem[\protect\citeauthoryear{Wei \& Zheng}{2012}]{Wei2012}	 Wei W. and  Zheng X.-P., 2012, Astroparticle Physics 37, 1 

\bibitem[\protect\citeauthoryear{Weissenborn et al.}{2011}]{Weissenborn2011} Weissenborn S., Sagert I., Pagliaria G., Hempel M.,  Schaffner-Bielich J., 2011 ApJ, 740, L14

\bibitem[\protect\citeauthoryear{Weissenborn et al.}{2012}]{Weissenborn2012}   Weissenborn S., Chatterjee D., Schaffner-Bielich J., 2012,  Nucl.Phys. A881, 62

\bibitem[\protect\citeauthoryear{Yoshida \& Kojima}{1997}]{Yoshida1997} Yoshida S.,  Kojima Y., 1997,  MNRAS, 289, 117

\bibitem[\protect\citeauthoryear{Yip et al.}{1999}]{YipChu1999}  Yip C. W.,  Chu M. C.,  Leung P. T., 1999, ApJ, 513, 849

\bibitem[\protect\citeauthoryear{Zdunik \& Haensel}{2013}]{Zdunik2013}  Zdunik J. L., Haensel P., 2013, A\&A 551, A61 

\end{thebibliography}
\end{document}